\documentclass[a4paper, oneside, twocolumn, notitlepage, 10pt]{extarticle_ecoc}
\usepackage{ecoc}

\usepackage{biblatex}
\begin{document}
\selectlanguage{english}    


\title{ Continuous-Variable Quantum Key Distribution Over 60 km Optical Fiber With Real Local Oscillator}%


\author{
    Adnan A.E. Hajomer\textsuperscript{*}, Hossein Mani,
    Nitin Jain, Hou-Man Chin, Ulrik L. Andersen, Tobias Gehring
}

\maketitle                  


\begin{strip}
 \begin{author_descr}

    Center for Macroscopic Quantum States (bigQ), Department of Physics, Technical University of Denmark, 2800 Kongens Lyngby, Denmark,
   \textsuperscript{*}\textcolor{blue}{\uline{ aaeha@dtu.dk}} 



 \end{author_descr}
\end{strip}

\setstretch{1.1}
\renewcommand\footnotemark{}
\renewcommand\footnoterule{}
\let\thefootnote\relax\footnotetext{978-1-6654-3868-1/21/\$31.00 \textcopyright 2021 IEEE}


\begin{strip}
  \begin{ecoc_abstract}
    We report the first continuous-variable quantum key distribution experiment that enables the generation of secure key over a 60 km fiber channel with locally generated local oscillator. This is achieved by controlling the excess noise using machine learning for phase noise compensation while operating the system at a low modulation variance.
  \end{ecoc_abstract}
\end{strip}


\section{Introduction}
Quantum key distribution (QKD) is an information-theoretically secure method to distribute secret keys between communication parties (Alice and Bob) based on the principles of quantum mechanics~\supercite{scarani2009security,pirandola2020advances}. Fundamentally, the secret key rate scales inversely with transmission distance~\supercite{pirandola2017fundamental} and as amplification of quantum states is incompatible with secret key generation, trusted or untrusted relays are required to achieve QKD over long distances. Naturally, it is desirable to extend the distance between relays as much as possible.


Since the first experiment of QKD over a 32-cm free space channel\supercite{bennett1992experimental}, considerable efforts have been devoted to perform QKD with channel lengths in the few 100s of km range\supercite{pirandola2020advances}. Most of these experiments employed so-called discrete-variable schemes that use single photon detectors, which are not standard telecom equipment and require special cooling units during operation. In contrast, encoding the secret key bit information in the phase and amplitude quadratures of the electro-magnetic light field and then decoding it with a coherent receiver in so-called continuous-variable (CV) QKD, offers the use of standard telecom components that work at room temperature\supercite{grosshans2002continuous, weedbrook2004quantum}. However, long-distance CVQKD has been limited due to two main factors. One is excess noise \supercite{lodewyck2005controlling}, mainly attributed to the laser phase noise, and the other is  limited  information reconciliation efficiency \supercite{leverrier2008multidimensional}. 

To avoid phase noise, several CVQKD experiments have used the so-called transmitted local oscillator (TLO) scheme \supercite{lodewyck2007quantum, jouguet2013experimental, wang201525, huang2016long, zhang2020long}, in which the transmitter/Alice prepares both the weak quantum signal and a strong local oscillator (LO) from the same laser and sends them to the receiver/Bob on the same optical fiber channel, to provide a stable phase reference. However, this implementation allows the eavesdropper/Eve to manipulate the LO, resulting in different possible attacks \supercite{ma2013local,jouguet2013preventing}. Moreover, to reduce in-band excess noise due to leakage from the strong TLO to the weak quantum signal, both polarization and time multiplexing are required, making the experimental realization of the system more complicated \supercite{qi2007experimental}. One way to deal with the security and implementation issues of TLO-based CVQKD is to use an independent laser at Bob to generate the LO. This  real local oscillator (RLO) or local local oscillator (LLO) scheme \supercite{qi2015generating,huang2015high,kleis2017continuous, chin2021machine, laudenbach2019pilot}, however, exhibits higher excess noise compared with TLO CVQKD because of the residual phase noise after the phase compensation procedure \supercite{marie2017self}. The maximum distance covered by a LLO-based CVQKD experiment has been 40 km \supercite{kleis2017continuous,laudenbach2019pilot}. 

Here, we report the longest (to our knowledge) distance experiment of CVQKD with LLO over 60 km fiber channel. This remarkable range is made possible by operating the system at low modulation variance of 1.8 shot noise units (SNU), in which phase noise is not a dominant factor. Besides, we use a machine learning (ML) framework for phase noise compensation so that a small residual phase noise can be maintained  at a low pilot power\supercite{chin2021machine}. Our system employs a continuous-wave (CW) laser and digital mode shaping, obviating the need of an additional amplitude modulator for pulse carving. As for error correction, we perform information reconciliation (IR) based on a multi-dimensional scheme using a multi-edge-type low-density-parity-check error correcting code \supercite{mani2021multiedge} with an efficiency of 94.31\% .
\begin{figure*}[t]
   \centering
    \includegraphics[width= 0.9\linewidth, keepaspectratio]{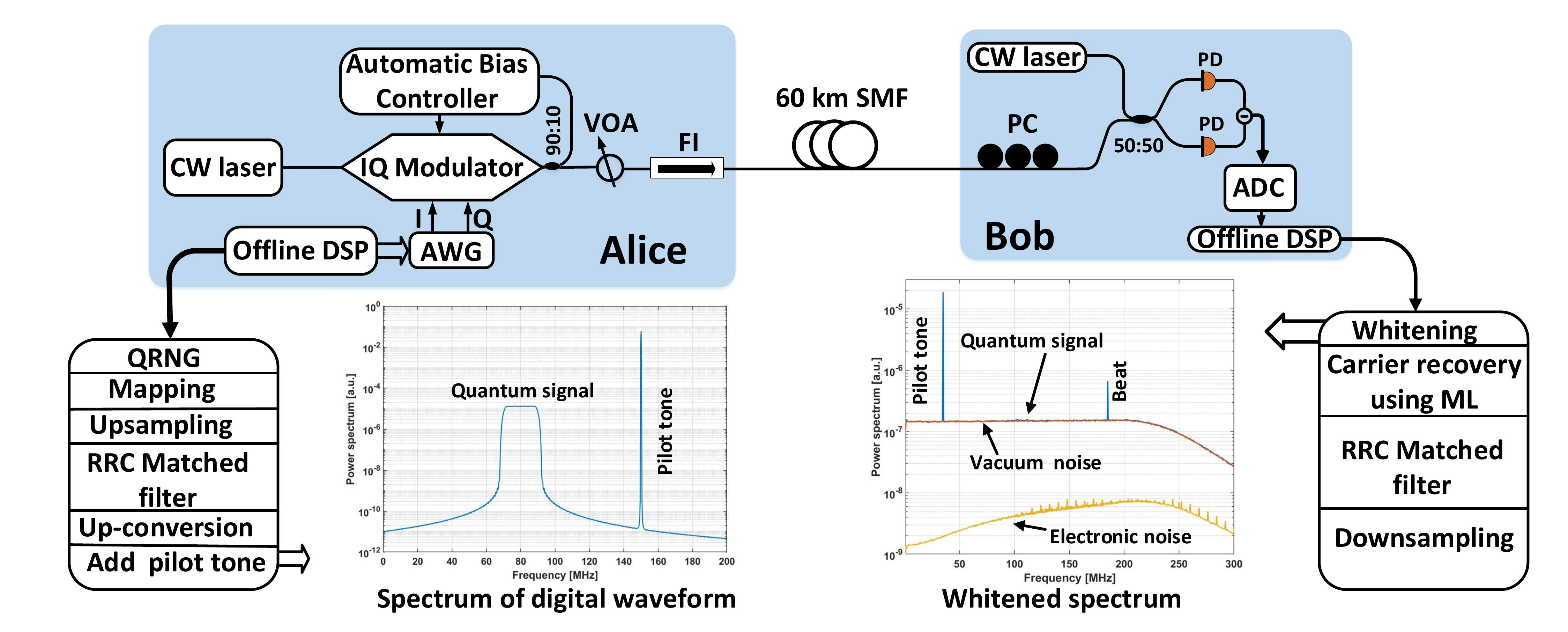}
    \caption{Experimental setup and DSP routine. QRNG: quantum random number generator; CW: continuous-wave laser; AWG: arbitrary waveform generator; VOA: variable optical attenuator; PC: polarization controller; ADC: analog-to-digital converter; PD: photodiode, FI: Faraday isolator, RRC: root raised cosine.}
    \label{fig1}
\end{figure*}
\vspace{-1em}
\section{Residual phase noise} 
To estimate the relative phase between Alice's and Bob's free running lasers in LLO CVQKD, pilot-aided techniques, in which a classical reference signal known as pilot tone is transmitted together with the quantum signal, have been used \supercite{qi2015generating,huang2015high,kleis2017continuous, chin2021machine}. However, the estimated  phase from these methods is not exactly equal to the actual relative phase of the quantum signal, which results in residual phase noise after phase compensation. This phase noise is the main contributor to the total excess noise that limits long-distance LLO CVQKD. In our case, where we use a Gaussian-modulated coherent state (GMCS) protocol, the phase noise can be expressed as \supercite{marie2017self},

\begin{equation}\label{eq:eq2}
\xi_{\text{phase}} = 2V_{\text{mod}}\left(1-e^{-\frac{V_{\text{est}}}{2}}\right),
\end{equation}
where $V_{\text{est}}$ is the variance of the residual phase, defined as the variance of the difference between the actual phase of the quantum signal and the estimated value, and $V_{\text{mod}}$ is the modulation variance. 

From equation  \ref{eq:eq2} it is clear that one can quantitatively reduce the phase noise either by reducing $V_{\text{est}}$,  for instance through better phase estimation, or by operating the system at a low modulation variance. While the quality of phase estimation depends on the signal-to-noise ratio of the pilot tone, the latter option requires careful optimization as the secret key rate has a $V_{\text{mod}}$ dependence even without phase noise. It is nevertheless practical and simple to realize as the modulation variance can be easily fine tuned. However, as a pilot tone with low power is desirable in CVQKD (to minimise the leakage from the pilot tone to the quantum signal), maintaining a constant residual phase, while the optical loss increases with distance, is not possible using traditional phase estimation techniques \supercite{qi2015generating,huang2015high}. As an alternative, ML has been shown to be an effective way to maintain a relatively constant residual phase over a wide-range of optical loss for a fixed input pilot power \supercite{chin2021machine}. We take the advantage of ML and operate in the low modulation variance regime to reduce the excess noise in our LLO CVQKD system, and therefore extend the channel distance. 
\begin{figure*}[t]
   \centering
    \includegraphics[width=0.94\linewidth]{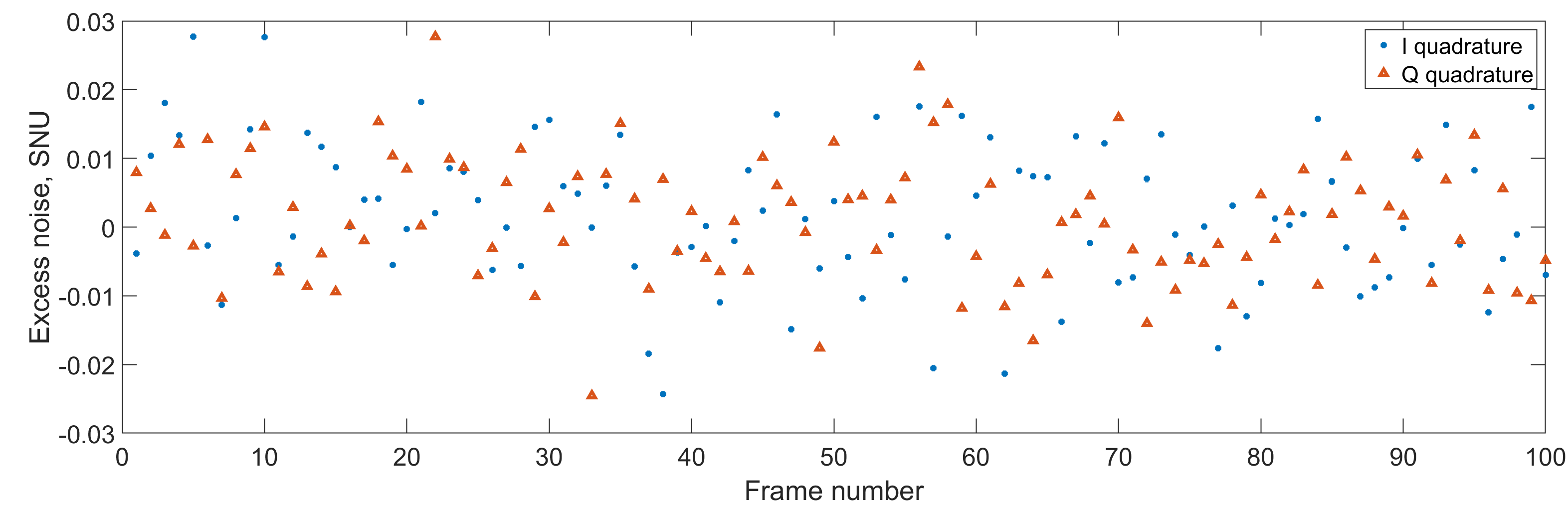}
    \caption{Measured excess noise.}
    \label{fig:figure2}
\end{figure*}
\vspace{-2em}
\section{Experimental setup}
Our experimental setup is shown in Fig.~\ref{fig1}. At Alice, a 20 Mbaud quantum signal was generated using offline digital signal processing (DSP). The transmitted symbols were drawn from a quantum number generator (QRNG) \supercite{Gehring2021qrng}. These symbols were upsampled to 1 Gsample/s and pulse-shaped by a root raised cosine (RRC) filter with a roll of factor of 0.2. The samples were then frequency shifted to 80 MHz and frequency multiplexed with a pilot tone at 150 MHz for carrier phase estimation. The spectrum of the digital waveform is shown in the inset of Fig.~\ref{fig1}.  The generated  digital waveforms were uploaded to an arbitrary waveform generator (AWG) with 16 bit resolution and sampling frequency of 1 GSample/s.  An in-phase and quadrature (IQ) modulator driven by the AWG was used to encode the ensemble of coherent states and the pilot tone onto a sideband of the optical carrier, generated from 1550 nm continuous-wave (CW) laser with $\approx $ 100 Hz linewidth. At the output of the IQ modulator, the quantum signal was attenuated using a variable optical attenuator (VOA), so that the modulation variance of the thermal state at the input of the 60 km single mode fiber (SMF) channel was 1.8 SNU. To avoid Trojan-horse attacks from the channel a Faraday isolator (FI) was added. 

At Bob, the polarization of the optical signal was adjusted using a manual polarization controller (PC). A radio frequency (RF) heterodyne detector, with a 3 dB bandwidth of 365 MHz, measured the optical signal after interference with the LLO on a 50:50 beam splitter. The LLO itself was generated from a CW laser with frequency shift of $\approx$ 180 MHz with respect to Alice's laser. The output of the detector was digitized using an analog-to-digital converter (ADC) with a sampling rate of 1 GSample/s, whose clock was synchronised together with that of the AWG to an external 10 MHz reference clock. The measurement time was divided into frames, each containing $10^7$  samples. To recover the transmitted symbols, offline DSP was applied to the recorded frames as follows: A whitening filter was first applied to the modulated signal, vacuum noise trace and electronic noise trace. The output of the whitening filter is depicted as an inset of Fig.~\ref{fig1}. The carrier phase estimation was performed using  an unscented Kalman filter on the pilot tone \supercite{chin2021machine}. Temporal synchronization was then achieved through a cross correlation between transmitted and received reference symbols, the RRC matched filtering and downsampling were applied to recover the quantum symbols. Finally, Alice and Bob perform IR and parameter estimation. 
\vspace{-1em}
\section{Results}
Fig.~\ref{fig:figure2} shows the measured excess noise variance at the output of the channel for 100 frames, each with $2 \times 10^5$ symbols. The average excess noise of \textit{I} and \textit{Q} quadrature is $1.1 \times 10^{-3}$ and $ 1.7 \times 10 ^{-3}$ SNU, respectively. The corresponding secret key rate fraction is computed in the asymptotic limit according to Ref.~\cite{chin2021machine}. Tab.~\ref{tab:table1} summarizes the experimental parameters used for secret key generation. Based on these parameters, we achieved a secret key faction of 0.0024 bits/symbol, corresponding to 0.0471 Mbits/s for a  symbol rate of 20 Mbaud. 
 
 Tab.~\ref{tab:table2} summarizes recent fiber based CVQKD demonstrations. For distances beyond 60 km, all demonstrations used TLO and pulse carving, in which an additional amplitude modulator is required. So far, the reported maximum distance of LLO CVQKD with CW laser was 40 km, however this system used a 8-state protocol\supercite{kleis2017continuous} instead of the GMCS protocol, which has a more mature security proof \supercite{weedbrook2004quantum}. 

\begin{table}[t]
   \centering
\caption{Experimental parameters. $\tau$: Trusted efficiency, $\eta$: Untrusted efficiency, $t$: trusted detection noise, $u$: untrusted channel noise, FER: frame error rate, $\beta$: IR efficiency.} \label{tab:table1}
\resizebox{1\hsize}{!}{
\begin{tabular}{|c|c|c|c|}
         \hline \bf Alice&\bf Bob & \bf Channel & \bf IR\\
         \hline  B = 20 MBaud & $\tau=0.68$  & $\eta=0.049$ & FER $=50\%$\\
         \hline  V$_{\text{mod}} = 1.8$ \text{SNU} & $t=58$ \text{mSNU} & $u= 1.3$ mSNU & $\beta = 94.31\%$\\
         \hline
\end{tabular}
}
\end{table}%

\begin{table}[t]
   \centering
\caption{ Summary of notable CVQKD demonstrations} \label{tab:table2}
\resizebox{1\hsize}{!}{
\begin{tabular}{|c|c|c|c|c|}
         \hline \bf Ref.  &\bf Pulsed/CW & \bf LO & \bf distance & \bf modulation\\
         \hline  \cite{lodewyck2007quantum} & Pulsed & TLO & 25 km & Gaussian\\
         \hline  \cite{wang201525} & Pulsed & TLO & 50 km & Gaussian\\
         \hline  \cite{jouguet2013experimental} & Pulsed & TLO & 80 km & Gaussian\\
         \hline  \cite{huang2016long} & Pulsed & TLO & 100 km & Gaussian\\
         \hline  \cite{zhang2020long} & Pulsed & TLO & 202.18 km & Gaussian\\
         
         \hline  \cite{qi2015generating,huang2015high} & Pulsed & LLO & 25 km & Gaussian\\
         \hline  \cite{laudenbach2019pilot} & Pulsed & LLO & 40 km &  Discrete \\
         \hline  \cite{kleis2017continuous} & CW & LLO & 40 km & Discrete \\
         \hline  \cite{chin2021machine} & CW & LLO & 20 km & Gaussian\\
         \hline  current work & CW & LLO & 60 km & Gaussian\\
         
         \hline
\end{tabular}
}
\end{table}

\section{Conclusions}
 We have reported a long-distance experiment that extends the security range of LLO CVQKD systems to record length of 60 km. This was made possible by taking advantage of a machine learning framework for phase noise compensation and operating the system with low modulation variance  to minimize system excess noise. This work is a step forward to close the gap between LLO- and TLO-CVQKD systems' performance, while maintaining a high level of security and lowering the implementation complexity.



\printbibliography

@article{pirandola2020advances,
  title={Advances in quantum cryptography},
  author={Pirandola, Stefano and Andersen, Ulrik L and Banchi, Leonardo and Berta, Mario and Bunandar, Darius and Colbeck, Roger and Englund, Dirk and Gehring, Tobias and Lupo, Cosmo and Ottaviani, Carlo and others},
  journal={Adv. Opt. Photonics},
  volume={12},
  number={4},
  pages={1012--1236},
  year={2020},
  publisher={Optical Society of America}
}

@article{pirandola2017fundamental,
  title={Fundamental limits of repeaterless quantum communications},
  author={Pirandola, Stefano and Laurenza, Riccardo and Ottaviani, Carlo and Banchi, Leonardo},
  journal={Nat. commun.},
  volume={8},
  number={1},
  pages={1--15},
  year={2017},
  publisher={Nature Publishing Group}
}

@article{scarani2009security,
  title={The security of practical quantum key distribution},
  author={Scarani, Valerio and Bechmann-Pasquinucci, Helle and Cerf, Nicolas J and Dusek, Miloslav and Lutkenhaus, Norbert and Peev, Momtchil},
  journal={Rev. Mod. Phys.},
  volume={81},
  number={3},
  pages={1301},
  year={2009},
  publisher={APS}
}

@article{bennett1992experimental,
  title={Experimental quantum cryptography},
  author={Bennett, Charles H and Bessette, Fran{\c{c}}ois and Brassard, Gilles and Salvail, Louis and Smolin, John},
  journal={J. cryptology},
  volume={5},
  number={1},
  pages={3--28},
  year={1992},
  publisher={Springer}
}

@article{grosshans2002continuous,
  title={Continuous variable quantum cryptography using coherent states},
  author={Grosshans, Fr{\'e}d{\'e}ric and Grangier, Philippe},
  journal={Phys. Rev. Lett.},
  volume={88},
  number={5},
  pages={057902},
  year={2002},
  publisher={APS}
}

@article{weedbrook2004quantum,
  title={Quantum cryptography without switching},
  author={Weedbrook, Christian and Lance, Andrew M and Bowen, Warwick P and Symul, Thomas and Ralph, Timothy C and Lam, Ping Koy},
  journal={Phys. Rev. Lett.},
  volume={93},
  number={17},
  pages={170504},
  year={2004},
  publisher={APS}
}

@article{leverrier2008multidimensional,
  title={Multidimensional reconciliation for a continuous-variable quantum key distribution},
  author={Leverrier, Anthony and All{\'e}aume, Romain and Boutros, Joseph and Z{\'e}mor, Gilles and Grangier, Philippe},
  journal={Phys. Rev. A},
  volume={77},
  number={4},
  pages={042325},
  year={2008},
  publisher={APS}
}

@article{lodewyck2005controlling,
  title={Controlling excess noise in fiber-optics continuous-variable quantum key distribution},
  author={Lodewyck, J{\'e}r{\^o}me and Debuisschert, Thierry and Tualle-Brouri, Rosa and Grangier, Philippe},
  journal={Phys. Rev. A},
  volume={72},
  number={5},
  pages={050303},
  year={2005},
  publisher={APS}
}

@article{lodewyck2007quantum,
  title={Quantum key distribution over 25 km with an all-fiber continuous-variable system},
  author={Lodewyck, J{\'e}r{\^o}me and Bloch, Matthieu and Garc{\'i}a-Patr{\'o}n, Ra{\'u}l and Fossier, Simon and Karpov, Evgueni and Diamanti, Eleni and Debuisschert, Thierry and Cerf, Nicolas J and Tualle-Brouri, Rosa and McLaughlin, Steven W and others},
  journal={Phys. Rev. A},
  volume={76},
  number={4},
  pages={042305},
  year={2007},
  publisher={APS}
}

@article{jouguet2013experimental,
  title={Experimental demonstration of long-distance continuous-variable quantum key distribution},
  author={Jouguet, Paul and Kunz-Jacques, S{\'e}bastien and Leverrier, Anthony and Grangier, Philippe and Diamanti, Eleni},
  journal={Nat. Photonics},
  volume={7},
  number={5},
  pages={378--381},
  year={2013},
  publisher={Nature Publishing Group}
}

@article{wang201525,
  title={25 MHz clock continuous-variable quantum key distribution system over 50 km fiber channel},
  author={Wang, Chao and Huang, Duan and Huang, Peng and Lin, Dakai and Peng, Jinye and Zeng, Guihua},
  journal={Scientific Reports},
  volume={5},
  number={1},
  pages={1--8},
  year={2015},
  publisher={Nature Publishing Group}
}

@article{huang2016long,
  title={Long-distance continuous-variable quantum key distribution by controlling excess noise},
  author={Huang, Duan and Huang, Peng and Lin, Dakai and Zeng, Guihua},
  journal={Scientific Reports},
  volume={6},
  number={1},
  pages={1--9},
  year={2016},
  publisher={Nature Publishing Group}
}

@article{zhang2020long,
  title={Long-distance continuous-variable quantum key distribution over 202.81 km of fiber},
  author={Zhang, Yichen and Chen, Ziyang and Pirandola, Stefano and Wang, Xiangyu and Zhou, Chao and Chu, Binjie and Zhao, Yijia and Xu, Bingjie and Yu, Song and Guo, Hong},
  journal={Phys. rev. lett.},
  volume={125},
  number={1},
  pages={010502},
  year={2020},
  publisher={APS}
}

@article{ma2013local,
  title={Local oscillator fluctuation opens a loophole for Eve in practical continuous-variable quantum-key-distribution systems},
  author={Ma, Xiang-Chun and Sun, Shi-Hai and Jiang, Mu-Sheng and Liang, Lin-Mei},
  journal={Phys. Rev. A},
  volume={88},
  number={2},
  pages={022339},
  year={2013},
  publisher={APS}
}

@article{jouguet2013preventing,
  title={Preventing calibration attacks on the local oscillator in continuous-variable quantum key distribution},
  author={Jouguet, Paul and Kunz-Jacques, S{\'e}bastien and Diamanti, Eleni},
  journal={Phys. Rev. A},
  volume={87},
  number={6},
  pages={062313},
  year={2013},
  publisher={APS}
}

@article{qi2007experimental,
  title={Experimental study on the Gaussian-modulated coherent-state quantum key distribution over standard telecommunication fibers},
  author={Qi, Bing and Huang, Lei-Lei and Qian, Li and Lo, Hoi-Kwong},
  journal={Phys. Rev. A},
  volume={76},
  number={5},
  pages={052323},
  year={2007},
  publisher={APS}
}

@article{qi2015generating,
  title={Generating the local oscillator “locally” in continuous-variable quantum key distribution based on coherent detection},
  author={Qi, Bing and Lougovski, Pavel and Pooser, Raphael and Grice, Warren and Bobrek, Miljko},
  journal={Phys. Rev. X},
  volume={5},
  number={4},
  pages={041009},
  year={2015},
  publisher={APS}
}

@article{huang2015high,
  title={High-speed continuous-variable quantum key distribution without sending a local oscillator},
  author={Huang, Duan and Huang, Peng and Lin, Dakai and Wang, Chao and Zeng, Guihua},
  journal={Optics lett.},
  volume={40},
  number={16},
  pages={3695--3698},
  year={2015},
  publisher={Optical Society of America}
}

@article{kleis2017continuous,
  title={Continuous variable quantum key distribution with a real local oscillator using simultaneous pilot signals},
  author={Kleis, Sebastian and Rueckmann, Max and Schaeffer, Christian G},
  journal={Opt. lett.},
  volume={42},
  number={8},
  pages={1588--1591},
  year={2017},
  publisher={Optical Society of America}
}

@article{chin2021machine,
  title={Machine learning aided carrier recovery in continuous-variable quantum key distribution},
  author={Chin, Hou-Man and Jain, Nitin and Zibar, Darko and Andersen, Ulrik L and Gehring, Tobias},
  journal={ npj Quantum Inf.},
  volume={7},
  number={1},
  pages={1--6},
  year={2021},
  publisher={Nature Publishing Group}
}

@article{marie2017self,
  title={Self-coherent phase reference sharing for continuous-variable quantum key distribution},
  author={Marie, Adrien and All{\'e}aume, Romain},
  journal={Phys. Rev. A},
  volume={95},
  number={1},
  pages={012316},
  year={2017},
  publisher={APS}
}

@article{mani2021multiedge,
  title={Multiedge-type low-density parity-check codes for continuous-variable quantum key distribution},
  author={Mani, Hossein and Gehring, Tobias and Grabenweger, Philipp and {\"O}mer, Bernhard and Pacher, Christoph and Andersen, Ulrik Lund},
  journal={Phys. Rev. A},
  volume={103},
  number={6},
  pages={062419},
  year={2021},
  publisher={APS}
}

@article{Gehring2021qrng,
abstract = {Quantum random number generators promise perfectly unpredictable random numbers. A popular approach to quantum random number generation is homodyne measurements of the vacuum state, the ground state of the electro-magnetic field. Here we experimentally implement such a quantum random number generator, and derive a security proof that considers quantum side-information instead of classical side-information only. Based on the assumptions of Gaussianity and stationarity of noise processes, our security analysis furthermore includes correlations between consecutive measurement outcomes due to finite detection bandwidth, as well as analog-to-digital converter imperfections. We characterize our experimental realization by bounding measured parameters of the stochastic model determining the min-entropy of the system's measurement outcomes, and we demonstrate a real-time generation rate of 2.9 Gbit/s. Our generator follows a trusted, device-dependent, approach. By treating side-information quantum mechanically an important restriction on adversaries is removed, which usually was reserved to semi-device-independent and device-independent schemes.},
author = {Gehring, Tobias and Lupo, Cosmo and Kordts, Arne and {Solar Nikolic}, Dino and Jain, Nitin and Rydberg, Tobias and Pedersen, Thomas B. and Pirandola, Stefano and Andersen, Ulrik L.},
%doi = {10.1038/s41467-020-20813-w},
file = {:C\:/Users/nitjai/research/MyPubs/2021 - Homodyne-based quantum random number generator at 2.9 Gbps secure against quantum side-information.pdf:pdf},
%issn = {20411723},
journal = {Nat, Commun.},
number = {1},
pages = {1--11},
publisher = {Springer US},
title = {{Homodyne-based quantum random number generator at 2.9 Gbps secure against quantum side-information}},
volume = {12},
year = {2021}
}

@article{laudenbach2019pilot,
  title={Pilot-assisted intradyne reception for high-speed continuous-variable quantum key distribution with true local oscillator},
  author={Laudenbach, Fabian and Schrenk, Bernhard and Pacher, Christoph and Hentschel, Michael and Fung, Chi-Hang Fred and Karinou, Fotini and Poppe, Andreas and Peev, Momtchil and H{\"u}bel, Hannes},
  journal={Quantum},
  volume={3},
  pages={193},
  year={2019},
  publisher={Verein zur F{\"o}rderung des Open Access Publizierens in den Quantenwissenschaften}
}
\end{document}